\documentclass[prb,aps, twocolumn, amsmath,amssymb,superscriptaddress]{revtex4}
\usepackage{float}
\usepackage{amsmath}
\usepackage{amssymb}
\usepackage{amsfonts}
\usepackage{euscript}
\usepackage{enumerate}
\usepackage{hhline}
\usepackage{pslatex}
\usepackage{tabularx}
\usepackage[usenames,dvipsnames]{xcolor}
\usepackage[colorlinks,linkcolor=red,anchorcolor=blue,citecolor=blue]{hyperref}

\usepackage{graphicx}
\usepackage{dcolumn}
\usepackage{bm}
\usepackage[sort&compress]{natbib}

\makeatletter

\newcommand{\mum}{ \,\mu\text{m}}
\renewcommand{\@biblabel}[1]{#1. }
\renewcommand{\@dotsep}{500}
\renewcommand{\@pnumwidth}{0em}
\renewcommand{\l@figure}[2]{
\@dottedtocline{1}{1.5em}{2em}{Figure #1}{}\vspace{15pt}}

\begin{document}

\title{Dual-Resonance Enhanced Quantum Light-Matter Interactions In Deterministically Coupled Quantum-Dot-Micopillars}

\author{Shunfa Liu}
\thanks{These authors contributed equally}
\affiliation{State Key Laboratory of Optoelectronic Materials and Technologies, School of Physics, School of Electronics and Information Technology, Sun Yat-sen University, Guangzhou 510275, China}

\author{Yuming Wei}
\thanks{These authors contributed equally}
\affiliation{State Key Laboratory of Optoelectronic Materials and Technologies, School of Physics, School of Electronics and Information Technology, Sun Yat-sen University, Guangzhou 510275, China}

\author{Xueshi Li}
\affiliation{State Key Laboratory of Optoelectronic Materials and Technologies, School of Physics, School of Electronics and Information Technology, Sun Yat-sen University, Guangzhou 510275, China}

\author{Ying Yu}
\thanks{yuying26@mail.sysu.edu.cn}
\affiliation{State Key Laboratory of Optoelectronic Materials and Technologies, School of Physics, School of Electronics and Information Technology, Sun Yat-sen University, Guangzhou 510275, China}

\author{Jin Liu}
\thanks{liujin23@mail.sysu.edu.cn}
\affiliation{State Key Laboratory of Optoelectronic Materials and Technologies, School of Physics, School of Electronics and Information Technology, Sun Yat-sen University, Guangzhou 510275, China}

\author{Siyuan Yu}
\affiliation{State Key Laboratory of Optoelectronic Materials and Technologies, School of Physics, School of Electronics and Information Technology, Sun Yat-sen University, Guangzhou 510275, China}
\affiliation{Photonics Group, Merchant Venturers School of Engineering, University of Bristol, Bristol BS8 1UB, United Kingdom}

\author{Xuehua Wang}
\affiliation{State Key Laboratory of Optoelectronic Materials and Technologies, School of Physics, School of Electronics and Information Technology, Sun Yat-sen University, Guangzhou 510275, China}

\date{\today}

\begin{abstract}
\noindent \textbf{Optical microcavities have widely been employed to enhance either the optical excitation or the photon emission processes for boosting light matter interactions at nanoscale. When both the excitation and emission processes are simultaneously facilitated by the optical resonances provided by the microcavities, as referred to the dual-resonance condition in this article, the performances of many nanophotonic devices approach to the optima. In this work, we present versatile accessing of dual-resonance conditions in deterministically coupled quantum-dot(QD)-micopillars, which enables emission from exciton (X) - charged exciton (CX) transition with improved single-photon purity. In addition, the rarely observed up-converted single-photon emission process is achieved under dual-resonance condition. We further exploit the vectorial nature of the high-order cavity modes to significantly improve the excitation efficiency under the dual-resonance condition. The dual-resonance enhanced light-matter interactions in the quantum regime provides a viable path for developing integrated quantum photonic devices based on cavity quantum electrodynamics (QED) effect e.g., highly-efficient quantum light sources and quantum logical gates.}
\end{abstract}

\maketitle

Last decade has witnessed significant advances in nanophotonics by harnessing the enhanced light-matter interaction in optical microcavities~\cite{Vahala2003}. E.g., cavity enhanced scattering and excitation enable the realization of biosensing with sensitivity down to single-molecule level~\cite{Vollmer2008,Baaske2014,Toropov2021} and highly efficient optical harmonic generations~\cite{LiuZJ2019,LiuHZ2018}. On the emission part, microcavities can modify the photonic environments of the nanoscale quantum emitters, resulting in faster radiative emission rate and better far-field radiation directionality~\cite{Gerard1998,Liu2018,Liu2019,Wang2019}. However, most of the nanophotonic devices based on high quality (Q) dielectric microcavities, to date, only involves the single resonance condition either for boosting the excitations or improving the photon emissions. Ideally, it is possible and highly desirable to simultaneously enhance both the excitation and the emission processes under the multiple-resonance condition, which is however technologically challenging especially for the dielectric microcavities with high-Q factors. Only until very recently, dual and even triply resonances conditions have been achieved in photonic crystal cavities, micro-rings and microspheres, which leads to the unprecedented device performances including Raman laser~\cite{Takahashi2013,Yu2020}, frequency conversion~\cite{Xue2017,Lu2019}, surface nonlinear optics~\cite{Zhang2019} and on-chip optical parametric oscillation with a record low threshold~\cite{Marty2021}. For single semiconductor QDs, most studies are focused on the enhancements of the emission process to pursue optimal single-photon sources~\cite{Somaschi2016Near,wang2019towards,tomm2021bright}. While the cavity enhanced P-shell excitation~\cite{Nomura2006}, wetting layer excitation~\cite{Kaniber2009} and phonon assisted excitation\cite{Madsen2014} have been observed by utilizing the high-order cavity modes of photonic nanocavities, the dual-resonance enhanced excitation-emission process has not been reported yet. In this work, we present versatile accessing of the dual-resonance conditions in deterministically coupled QD-micopillars operating in the cavity QED regime. By carefully engineering the fundamental mode and the high-order mode of the micropillars, we have realized both up-converted and down-converted single-photon emission under the dual-resonance condition. In particular, the intra-dot transitions between the  X and the CX in the down-conversion process effectively suppress the carrier recapturing process by the defects states in semiconductor and therefore improve the single-photon purity of the emission. We further show that the excitation efficiency under dual-resonance condition can be greatly improved by utilizing the vectorial excitation beams with the same polarization states as the high-order cavity modes\cite{Fang2017,LeKien2017}.

\begin{center}
	\begin{figure*}
		\begin{center}
			\includegraphics[width=0.85\linewidth]{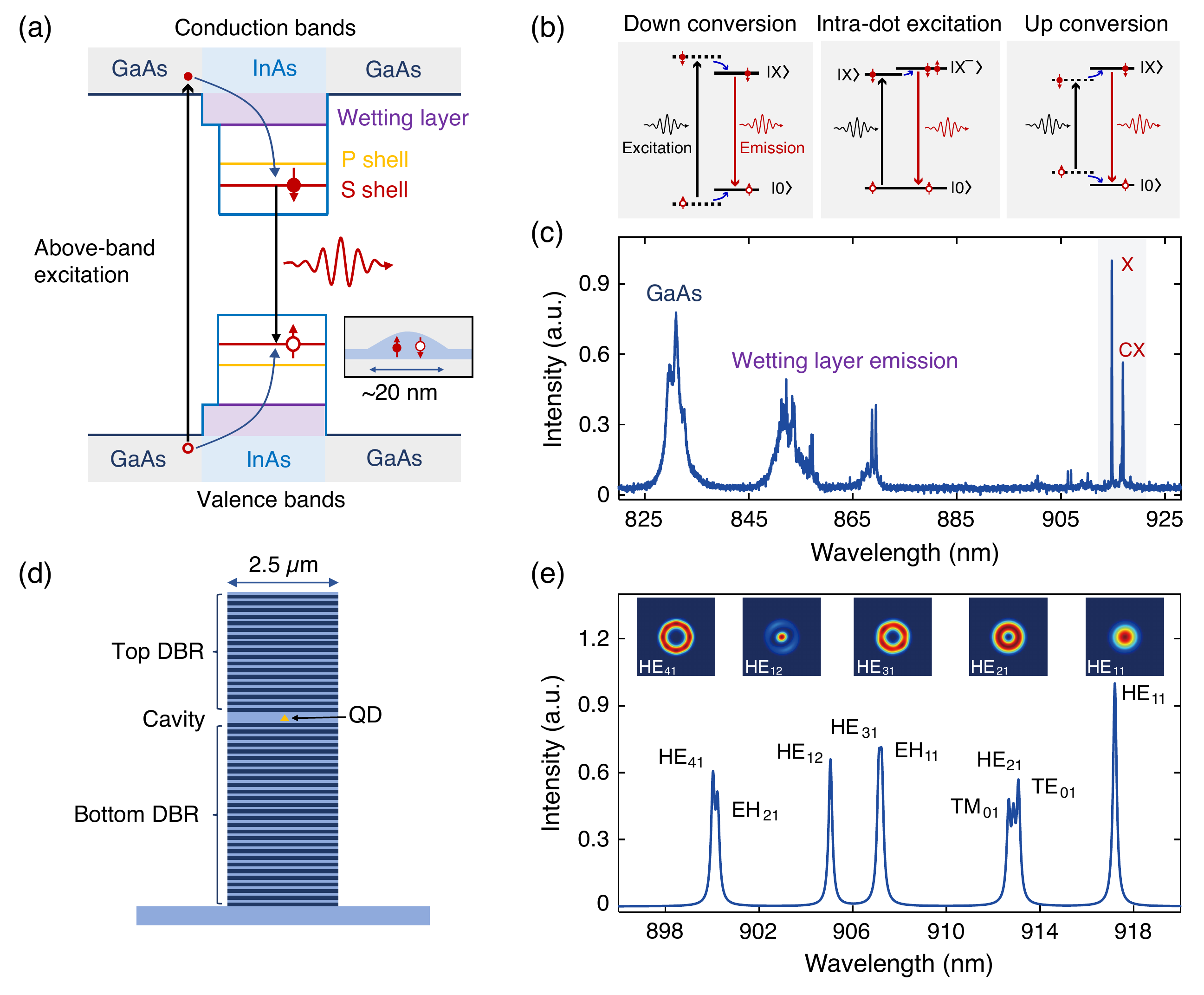}
			\caption{\textbf{QD-micropillar system.}  (a) Schematic of the energy diagram of InAs QD embedded in GaAs matrix. The quantum confinement of carriers in the QD results in discrete energy levels. (b) Different excitation schemes for triggering the single-photon emissions from exciton states in QDs used in the work. All the excitation methods can be improved by the optical resonances provided from microcavities. The black and red lines represent the optical excitation and photon emission processes respectively while the blue lines denote the non-radiative process. (c) Representative spectrum of a single QD under the low-power above-band excitation. (d) Schematic of the coupled QD-micropillar system. (e) The calculated radiation spectrum and the mode profiles of the micropillar cavity with a diameter of 2.5 $\mu$m.}
			\label{fig:Fig1}
		\end{center}
	\end{figure*}
\end{center}

We use single InAs QDs embedded in a GaAs matrix as quantum emitters\cite{1lodahl2015interfacing,Buckley2012}, as schematically shown in Fig.~\ref{fig:Fig1}(a). Due to the quantum confinement of carriers at nanoscale, the QD exhibit atomic-like discrete energy levels, such as S-shell and P-shell. Carriers can be excited by using a laser with the energy higher than the bandgap of GaAs, referred as to above-band excitation (denoted by the thick black arrow). The created carriers in the GaAs material then relax to the lowest excited states of the QD via electron-phonon scattering before the radiative recombination process of single-photon emissions. The longitudinal optical or acoustic phonons in the solid-state provide additional degree of freedom over the atomic systems to excite the QDs via both down- and up- conversion processes\cite{Englund2010,Quilter2015,Reindl2019,Pooley2012}, as shown in Fig.~\ref{fig:Fig1}(b). More interestingly, the transition between exciton states with different charge configurations (referred as to intra-dot excitation in this work) can also be utilized to trigger the radative process as recently demonstrated in two-dimensional semiconductor~\cite{Jones2016}. A representative emission spectrum of a singe InAs QD under the low-power above-band excitation is presented in Fig.~\ref{fig:Fig1}(c), exhibiting a broad band GaAs band edge emission, a wetting layer emission and sharp X and CX lines. To build a coupled QD-micropillar system, the single InAs QD is embedded in the centre of a semiconductor planar cavity consisting of a $\lambda$-thick GaAs spacer sandwiched by  GaAs/Al$_{0.9}$Ga$_{0.1}$As distributed Bragg reflectors (DBR) with 18(26) top (bottom) pairs grown via molecular beam epitaxy. Micropillars are then fabricated from the planar cavity in order to reduce the cavity mode volume for further enhancing photon-exciton interaction, as schematically shown in Fig.~\ref{fig:Fig1}(d). Micropillar supports a series of cavity modes with sharp resonances over a broad bandwidth~\cite{Gerard1996,Reitzenstein2010}. In Fig.~\ref{fig:Fig1}(e), the mode family of a 2.5 $\mum$ micropillar with the planar cavity resonance at 920 nm is calculated by the finite difference time domain (FDTD) simulation with the insets representing the intensity profiles of a few representative  modes. Among all the cavity modes, the fundamental mode HE$_{11}$ exhibits the highest Q-factor, lowest mode volume and near Gaussian far-field pattern, which is widely used for building high-performance single-photon sources via the cavity QED effect~\cite{Somaschi2016Near,Ding2016demand,He2016,Su2018}.

\begin{figure*}
	\includegraphics[width=0.9\linewidth]{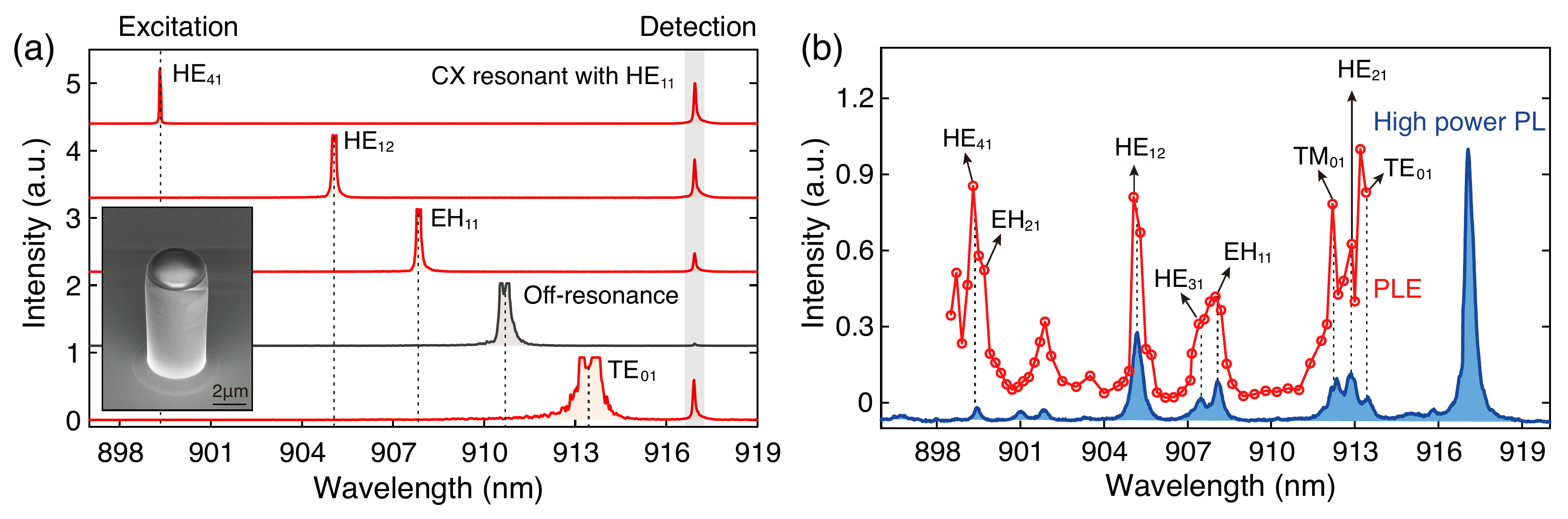}
	\caption{\textbf{Identification of the cavity modes of the micropillar.}  (a) PL spectra for the QD-micropillar excited by a laser with varied wavelengths. The CX emission is resonant and therefore enhanced by the HE$_{11}$ mode. Appreciable CX emission is observed when excitation laser is resonant with one of the high-order cavity modes (red spectrum). In the off-resonance condition, the CX emission intensity is neglecgible. (black spectrum) Inset, SEM image of the fabricated micropillar. (b). PLE spectrum (red) for the CX state resonant with the HE$_{11}$ mode and the PL spectrum (blue) of the QD-cavity system under high-power above-band excitation condition. The high-power PL is shifted slightly in y direction for clarity.}
	\label{fig:Fig2}
\end{figure*}

\begin{figure*}
	\begin{center}
		\includegraphics[width=0.85\linewidth]{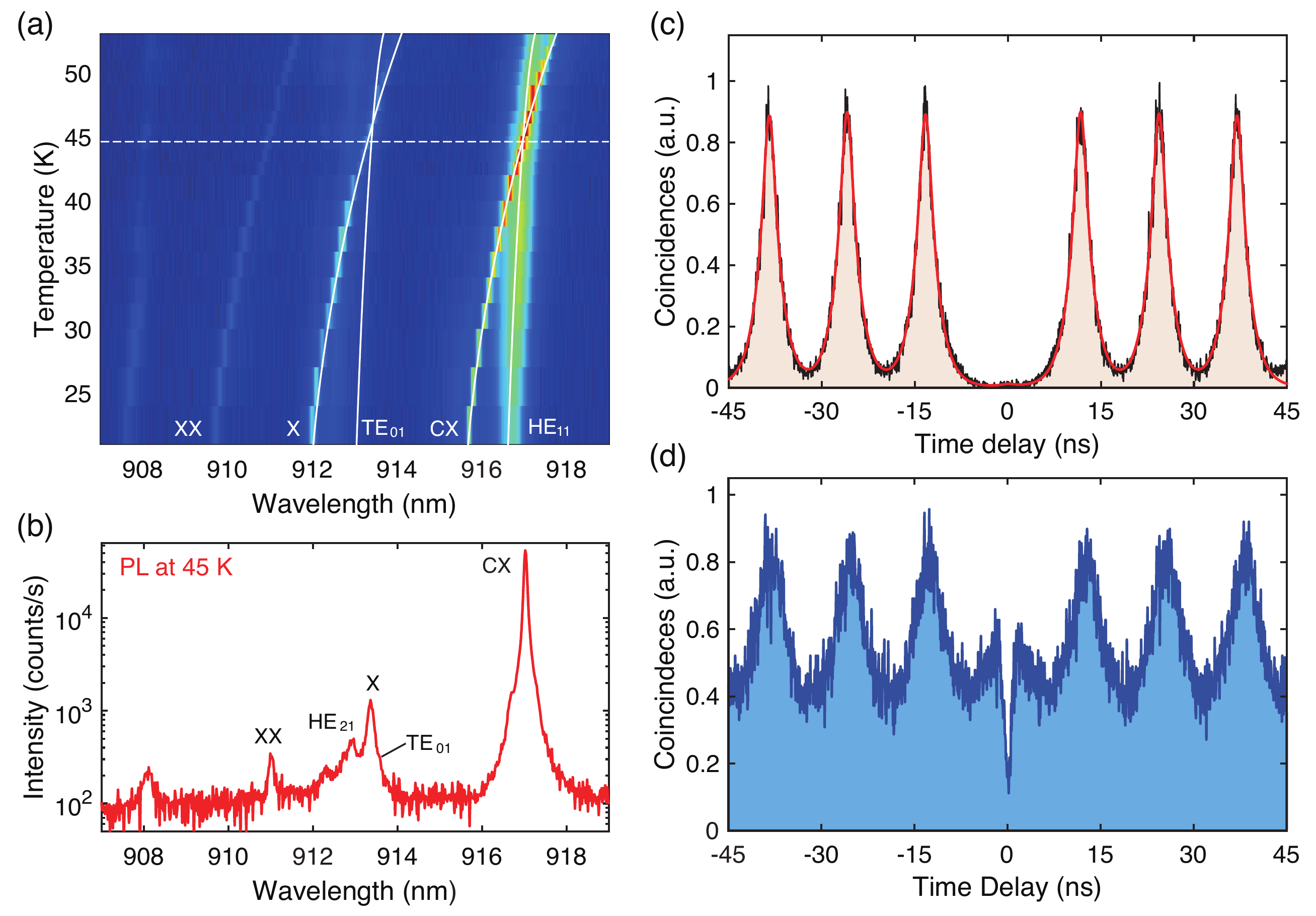}
		\caption{\textbf{Dual-resonance enhanced X-CX transition for the highly pure single-photon emission}. (a) Temperature dependent PL mapping of a QD coupled to a 2.5 $\mu$m diameter micropillar cavity, the white line is guide for the eyes. (b) Log scaled line cut of PL mapping in (a) from a QD obtained under above-band (780 nm) excitation at temperature of 45 K. At this temperature, the CX is tuned into resonance of the HE$_{11}$ mode and X is resonant with TE$_{01}$ mode.  (c) and (d) Hanbury Brown and Twiss (HBT) measurements of single-photon purity for dual-resonance enhanced the intra-dot excitation and the above-band excitation. Strong suppression of the carrier recapturing process is observed via intra-dot excitation, leading to a background-free $g^{(2)}(0)$ value as low as 0.01.}
		\label{fig:Fig3}
	\end{center}
\end{figure*}

The deterministically coupled QD-micropillars are fabricated by using the fluorescence imaging technique~\cite{He2016,Liu2017a,Liu2017} which ensures that the single QDs are both spectrally and spatially matched with the fundamental cavity mode (HE$_{11}$) of the micropillar. We first identify the mode family of the micropillar by scanning the excitation laser through all the high-order cavity modes and monitoring the emission from the CX state in resonance with the fundamental cavity mode (HE$_{11}$). The power of excitation laser is kept as a constant level of 0.5 mW before the objective lens to avoid saturating the QD. As along as the excitation laser is tuned to one of the high-order cavity modes, bright CX emission is observed (red spectra). On the contrary, the CX emission is barely detectable when the excitation laser is detuned from any of the high-order cavity modes, as shown by the black spectrum in Fig.~\ref{fig:Fig2}(a). The identifications of the high-order cavity modes under the dual-resonance condition is further quantified via photoluminescence excitation (PLE) spectrum (red points) in Fig.~\ref{fig:Fig2}(b) in which the emission intensity of the CX state is plotted as a function of the wavelength of the excitation laser. The high-order cavity modes are confirmed by another independent experiment in which cavity modes are mapped from the PL spectrum (blue curve) under high-power above-band excitation. In such a scenario, the single QD serves as a broadband internal light sources to efficiently probe all the cavity modes~\cite{Smolka2011}.  

 \begin{center}
	\begin{figure}
		\begin{center}
			\includegraphics[width=\linewidth]{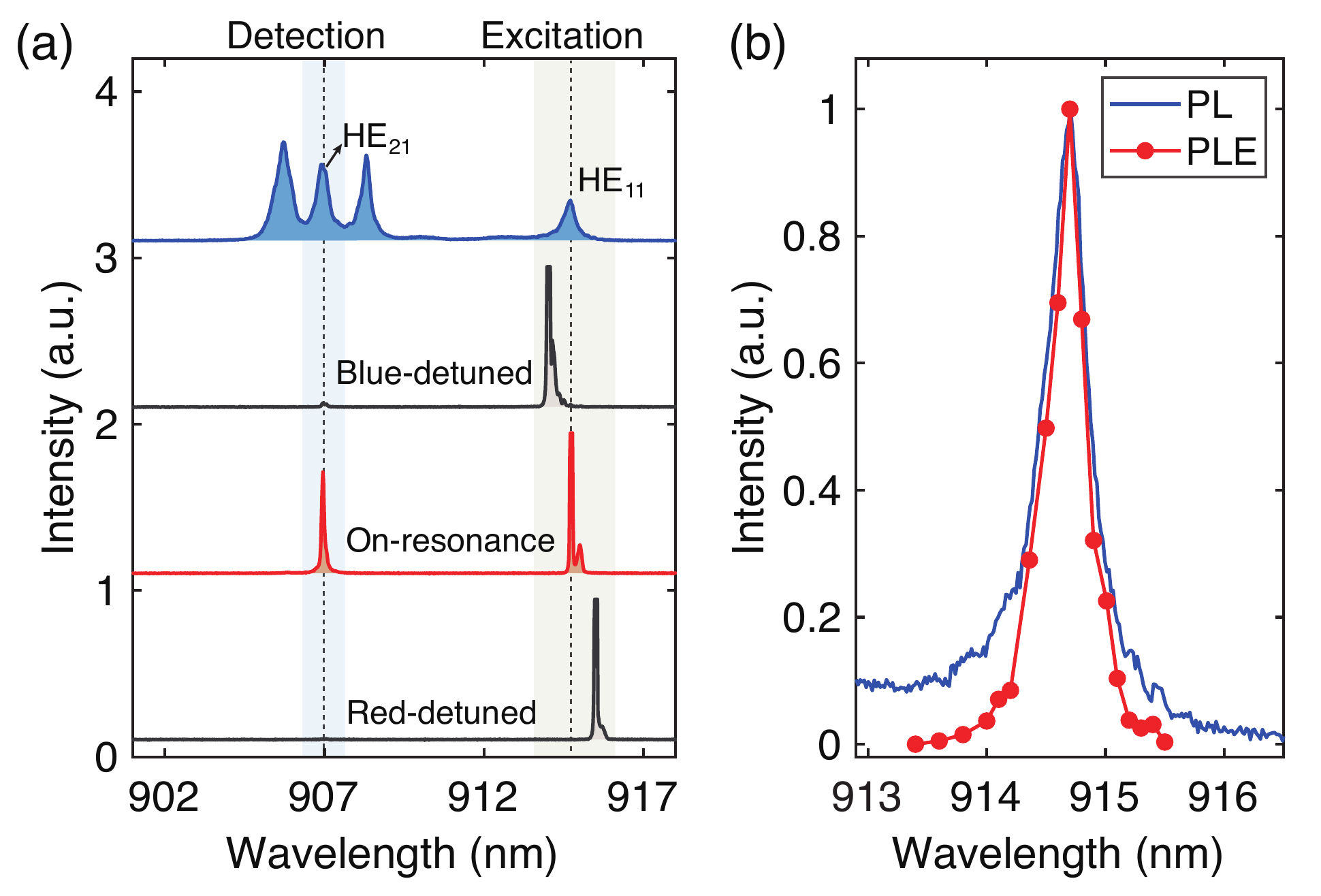}
			\caption{\textbf{Dual resonances enhanced up-converted excitation.}  (a) High-power above-band PL spectrum of a pillar with diameter of 2 $\mu$m (blue) and low-power PL spectra of a QD excited under dual resonances (red) and single resonance conditions (black). (b) Normalized high-power PL spectrum of HE$_{11}$ mode (blue) and PLE spectrum of the QD acquired by sweeping the excitation laser through HE$_{11}$ mode and detecting the QD emission intensity (red), the power of excitation laser is kept constant for PLE measurement.}
			\label{fig:Fig4}
		\end{center}
	\end{figure}
\end{center}

 We further show, for the first time, that the intra-dot transition process between the exciton states in S-shell of a QD can be facilitated under the dual-resonance condition, as recently demonstrated in two-dimensional semiconductor~\cite{Jones2016}. Such a process enables the high-purity single-photon emission due to the absence of carrier recapturing process by the defect state in the semiconductor. Due to the Coulomb interactions of the confined carriers, there is a slightly energy shift between the X and CX~\cite{Regelman2002,Ediger2007}. The intra-dot transition is implemented by tuning the excitation laser to match the energy of X state (913.4 nm) and monitoring the emission from the CX state (917.02 nm). The resonance condition can be reached simultaneously for both X and CX by tuning the temperature of the sample, as shown by the temperature dependent spectra in Fig.~\ref{fig:Fig3}(a). At 45 K, the X and CX states are simultaneously resonant to the TE$_{01}$ and HE$_{11}$ modes respectively, as presented in Fig.~\ref{fig:Fig3}(b). The photon statistics of the CX emission is examined by the Hanbury-Brown-Twiss (HBT) interferometer and the coincidence histogram of the second-order correlation function is presented in Fig.~\ref{fig:Fig3}(c). Under the dual-resonance enhanced intra-dot excitation, the coincidence event at the zero delay is almost vanishing with a near-zero background across the whole histogram, indicating the generation of single-photon emission with high-purity from the CX state. For comparison, the photon statistics of the CX emission under the above-band excitation condition (pumping at 780 nm) is presented in Fig.~\ref{fig:Fig3}(d), which shows significant background due to the carrier recapturing process by the defect states in the semiconductor~\cite{Dalgarno2008,Aichele2004,Yang2020}.

\begin{center}
	\begin{figure}
		\begin{center}
			\includegraphics[width=0.8\linewidth]{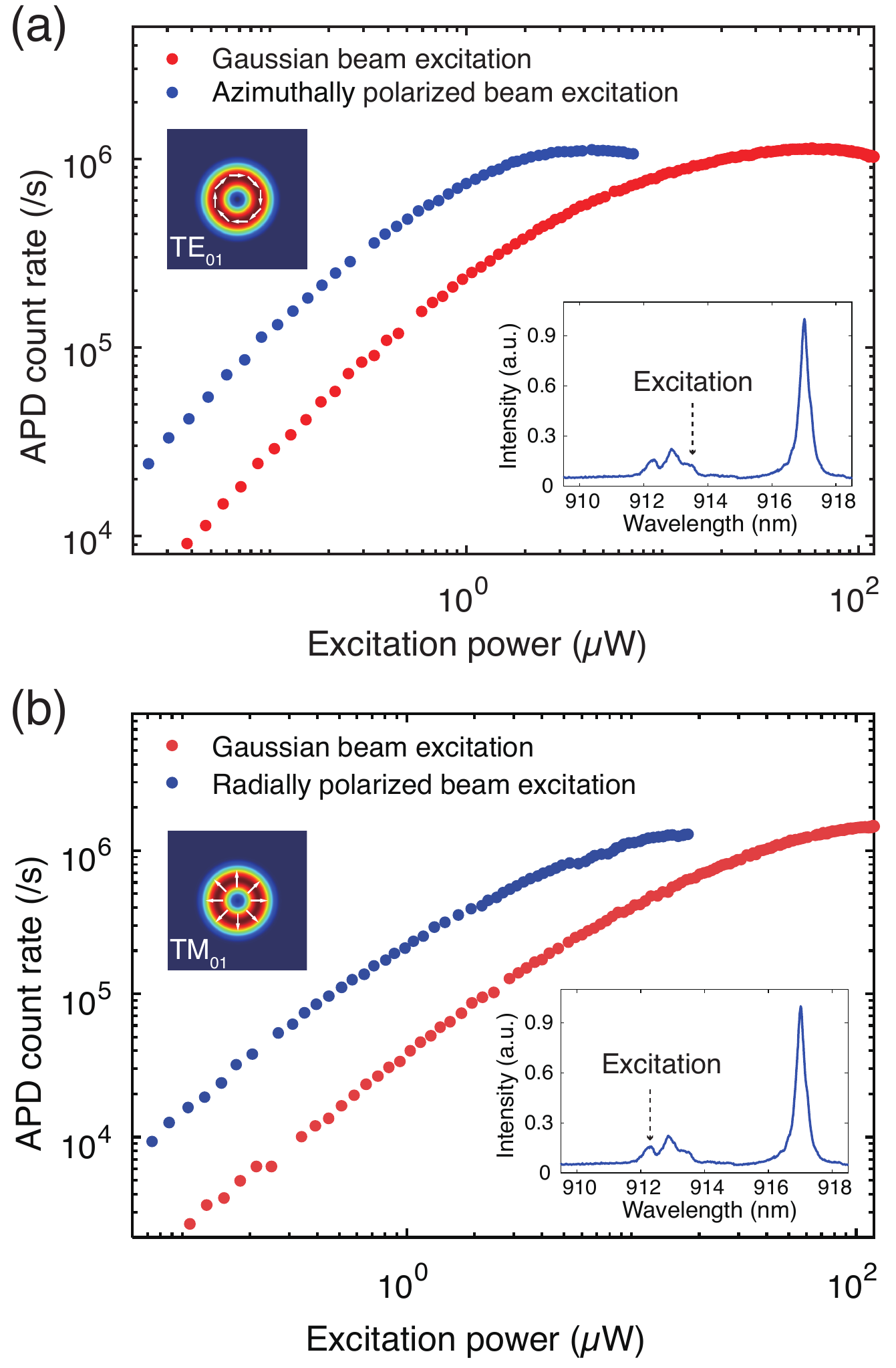}
			\caption{\textbf{Power dependence of QD emission excited by vectorial beams.} The CX is resonant with HE$_{11}$ mode, and excited via the azimuthally polarized TE$_{01}$ mode (blue) (a) and the radially polarized TM$_{01}$ mode (blue) (b) to match the polarization states of the high-order cavity modes. For comparison, a linear polarized Gaussian beam excitation (red) is implemented.}
			\label{fig:Fig5}
		\end{center}
	\end{figure}
\end{center}

Comparing to the down-conversion process, the up-conversion process is more challenging since it extracts energy out of the system. In the up-conversion process, one low-energy photon in the excitation laser absorb an acoustic phonon and results in the emission of a single photon from the exciton state. Such processes have been recently employed to cool the mechanical motions of a micro-resonator to its quantum ground state~\cite{Chan2011,Verhagen2012} or bulk temperature of semiconductors~\cite{Zhang2013,Zhang2016,Ha2016}, showing great potential in exploring fundamental quantum physics and exploiting novel optical refrigeration method for nanophotonic devices. As opposed to the dual-resonance enhanced down-conversion process, the up-conversion process utilizes the fundamental mode to boost the optical excitation and the high-order cavity modes to enhance the photon emission. As shown in Fig.~\ref{fig:Fig4}(a), the PL spectra under detuning conditions are presented. Under the high-power above-band excitation condition (blue spectrum), the cavity modes of HE$_{11}$, TE$_{01}$, HE$_{21}$ and TM$_{01}$ are clearly identified. The excitation laser is then scanned across the HE$_{11}$ mode to excite the X state resonant with the HE$_{21}$ mode. Under the dual-resonance condition (red spectrum), bright X state emission is observed. Such emission is nearly vanishing once the laser is either slightly red or blue-detuned from the HE$_{11}$ mode (Black Spectra). The PLE spectrum of the HE$_{11}$ mode matches excellently with the cavity resonance observed in the high-power PL spectrum under above-band excitation condition as shown in Fig.~\ref{fig:Fig4}(b), indicating that the up-conversion process is enhanced by the cavity. Further developments along this direction may result in the realizations of reduced electron-phonon interactions in the system and even the development of optical refrigeration for single QDs.

Finally, we show that the efficiency of the down-conversion process under the dual resonances condition can be further improved by engineering the polarization state of the excitation beam~\cite{Koshelev2020}. Instead of the linearly-polarized HE$_{11}$ mode, the  high-order cavity modes exhibit vectorial polarizations, e.g., the TE$_{01}$ mode is azimuthally polarized while TM$_{01}$ mode is radially polarized. Fig.~\ref{fig:Fig5}(a) shows the single-photon emission intensity of single photons from the CX state as a function of excitation laser power at 913.5 nm (resonant with the TE$_{01}$ mode) with different states of polarization. While the emission intensity at the saturation power is the same, the excitation power required to saturate the QD is reduced from 50 $\mu$W to 3 $\mu$W by switching the excitation laser from a linearly polarized Gaussian beam to a radially polarized vortex beam. Similar behavior can be observed for the excitation via the TM$_{01}$ mode, as shown in Fig.~\ref{fig:Fig5}(b), in which the saturation powers is reduced by a factor of 7 by using the azimuthally polarized excitation beam.

To conclude, we show versatile accessing of dual-resonance conditions for enhancing the light matter interactions in QD-micropillar devices operating in the cavity QED regime. The cavity mode family are independently identified in both the PLE measurement and the high-power PL spectrum under the above-band excitation condition. By exploiting the intra-dot excitation under the dual-resonance condition, the singe-photon purity of emitted photons is greatly improved compared to the above-band excitation condition due to the suppression of the carrier recapture process by the defects in the semiconductor. The up-converted emission is further demonstrated by using excitation via fundamental cavity mode and emission at the high-order cavity resonance. Such a process could be used to engineer electron- phonon interactions and optical refrigeration of single QDs. By engineering the polarization state of the excitation laser beam, the excitation efficiency can be further boosted. The QD-micropillar system under dual-resonance condition may serve as an ideal platform in solid state for investigating light matter interaction in the quantum regime and developing integrated quantum photonic devices with high-performances.


\noindent \textbf{Acknowledgements}
The authors wish to thank Lin Liu and Li-Dan Zhou for technical assistance in microfabrication. This research was supported by National
Key R\&D Program of China (2018YFA0306100, 2018YFA0306302), the National Natural Science Foundation of China (11874437, 61935009), Guangzhou Science and Technology Project (201805010004), the Natural Science Foundation of Guangdong (2018B030311027), the national super-computer center in Guangzhou, the National Natural Science Foundation of China (12074442, 91836303), and the Local Innovative and Research Teams Project of Guangdong Pearl River Talents Program (2017BT01X121).


\end{document}